\documentclass[amsmath,amssymb,12pt,superscriptaddress,nofootinbib]{revtex4}

\usepackage{graphicx}
\usepackage{dcolumn}
\usepackage{bm}
\usepackage{color}
\usepackage{enumitem}
  \usepackage{tabularx}
   \newcolumntype{C}{>{\centering\arraybackslash}X}
   \newcolumntype{L}{>{\raggedright\arraybackslash}X}
   \newcolumntype{R}{>{\raggedleft\arraybackslash}X}
\setlistdepth{10}
\usepackage[normalem]{ulem}

\newcommand{\dd}{\mathrm{d}}

\newcommand{\del}{\partial}

\definecolor{DarkBlue}{rgb}{0,0,0.7} 

\definecolor{DarkRed}{rgb}{0.65,0,0}

\begin{document}
\baselineskip5.5mm
%\lineskip5.5mm
%\date{\today}
%\if0

{\baselineskip0pt
\small
\leftline{\baselineskip16pt\sl\vbox to0pt{
%               \hbox{\it Division of Particle and Astrophysical Science, Nagoya University}
%%              \hbox{\it Instituto Superior T\'ecnico}
%%              \hbox{\it Nagoya University}
 %              \hbox{\it Department of Physics, Rikkyo University} 
%             \hbox{\it Rikkyo University}
%		\hbox{\it Yukawa Institute for Theoretical Physics, Waseda University} 
%               \hbox{\it Advanced Research Institute for Science and Engineering, Waseda University} 
%             \hbox{\it Waseda University}
                             \vss}}
\rightline{\baselineskip16pt\rm\vbox to20pt{
 		%\hbox{OCU-PHYS-xxx}
%            \hbox{AP-GR-xx}
\vspace{-1.5cm}
            \hbox{YITP-21-161}
           \hbox{RUP-21-23}
  %          \hbox{KEK-TH-2245}
\vss}}
}

\author{Chul-Moon~Yoo}\email{yoo@gravity.phys.nagoya-u.ac.jp}

\affiliation{
\fontsize{11pt}{10pt}\selectfont
Division of Particle and Astrophysical Science,
Graduate School of Science, \\Nagoya University, 
Nagoya 464-8602, Japan
\vspace{1.5mm}
}

\author{Tomohiro~Harada}\email{harada@rikkyo.ac.jp}
\affiliation{~
\fontsize{11pt}{10pt}\selectfont
Department of Physics, Rikkyo University, Toshima,
Tokyo 171-8501, Japan
\vspace{1.5mm}
}

\author{Shin'ichi~Hirano}\email{hirano.shinichi@a.mbox.nagoya-u.ac.jp}
\affiliation{
\fontsize{11pt}{10pt}\selectfont
Division of Particle and Astrophysical Science,
Graduate School of Science, \\Nagoya University, 
Nagoya 464-8602, Japan
\vspace{1.5mm}
}

\author{Hirotada~Okawa}\email{h.okawa@aoni.waseda.jp}

\affiliation{
\fontsize{11pt}{10pt}\selectfont
Waseda Institute for Advanced Study(WIAS), 
Waseda University, Shinjuku, Tokyo 169-0051, Japan
\vspace{1.5mm}
}

\author{Misao~Sasaki}\email{misao.sasaki@ipmu.jp}

\affiliation{
\fontsize{11pt}{10pt}\selectfont
Kavli Institute for the Physics and Mathematics of the Universe (WPI),\\
University of Tokyo, Kashiwa 277-8583, Japan
\vspace{1.5mm}
}

\affiliation{
\fontsize{11pt}{10pt}\selectfont
Center for Gravitational Physics, Yukawa Institute for Theoretical Physics, 
\\ Kyoto University, Kyoto 606-8502, Japan
\vspace{1.5mm}
}

\affiliation{
\fontsize{11pt}{10pt}\selectfont
Leung Center for Cosmology and Particle Astrophysics, \\
National Taiwan University, Taipei 10617, Taiwan 
\vspace{1.5mm}
}

\vskip-1.cm
\title{Primordial black hole formation from massless scalar isocurvature}

%\if0

\begin{abstract}
\baselineskip5.5mm 
We numerically study the primordial black hole (PBH) formation by an isocurvature
perturbation of a massless scalar field on super Hubble scales 
in the radiation-dominated universe. 
As a first step we perform simulations of spherically symmetric configurations.
For the initial condition, we employ the spatial gradient expansion and 
provide the general form of the growing mode solutions valid up through the
second order in this expansion.
The initial scalar field profile is assumed to be
Gaussian with a characteristic comoving wavenumber $k$; $\sim\exp(-k^2R^2)$, where 
$R$ is the radial coordinate.
We find that a PBH is formed for a sufficiently large amplitude of the scalar field profile.
Nevertheless, we find that the late time behavior of the gravitational collapse is 
dominated by the dynamics of the fluid but not by the scalar field,
which is analogous to the PBH formation from an adiabatic perturbation in the radiation-dominated universe. 
\end{abstract}

%\fi
%\baselineskip5.5mm

\maketitle
%\thispagestyle{empty}
%\pagebreak

%%%%%%%%%%%%%%%%%%%%%%%%%%%%%%%%%%%%%%%%%%%%%%%%%%%%%%%%%%%%%%%%
\section{Introduction}
%%%%%%%%%%%%%%%%%%%%%%%%%%%%%%%%%%%%%%%%%%%%%%%%%%%%%%%%%%%%%%%%

Recent technological advances opened the era of precision cosmology. 
The resolution which we can access through observations 
is getting finer and finer, and we now have a fairly accurate picture of the
primordial universe based on the inflationary universe scenario. 
However, the accessible range of scales through cosmological observations 
such as the cosmic microwave background and the large scale structure 
is very limited, and only little is known about the primordial state of the universe 
on small scales, say, on comoving scales smaller than the galactic scale.

In this situation, recently primordial black holes (PBHs) have been attracting
 a lot of attention. 
The possible existence of PBHs was suggested about half a century ago~\cite{Hawking:1971ei,Carr:1974nx}. 
There has been a renewed interest in PBHs 
thanks in particular to the progress in gravitational wave astrophysics/cosmology
(e.g., see Refs.~\cite{Carr:2020gox,Carr:2020xqk} for reviews on PBHs). 
Theoretically, many models which produce a substantial amount of PBHs have been
proposed in the context of the inflationary universe, or in alternative scenarios.

In most of these models, the PBH formation is attributed to the curvature 
perturbation generated during inflation. 
In other words, only the single adiabatic 
degree of freedom has been considered, while isocurvature degrees of freedom which
may play important roles in a universe consisting of multi-fluids have been
largely ignored. Only quite recently, the PBH formation from isocurvature 
perturbations of the cold dark matter was considered~\cite{Passaglia:2021jla}
(see also Ref.~\cite{Domenech:2021and} for the possible detection of gravitational waves from large dark matter isocurvature).

In the case of the conventional PBH formation scenario in the radiation-dominated 
universe, numerical simulations of the PBH formation are one of the crucial 
elements to obtain accurate theoretical 
predictions~(see e.g. \cite{Shibata:1999zs,Niemeyer:1999ak,Musco:2004ak,Polnarev:2006aa,Musco:2008hv,Polnarev:2012bi,Nakama:2013ica,Musco:2018rwt,Escriva:2019nsa,Escriva:2019phb,Yoo:2020lmg}). 
It is also certainly true for the case of PBH formation 
from isocurvature perturbations.
In this paper, we make a first attempt to perform such simulations by
considering an isocurvature perturbation of a massless scalar field 
in the radiation-dominated universe. We assume spherical symmetry and consider
the Gaussian spatial profile, $\phi(R)=\mu\exp(-k^2R^2/6)$, 
as the initial scalar field configuration, where $\mu$ is the amplitude, $k$ is 
the characteristic wavenumber of the perturbation, and $R$ is the 
radial coordinate.

To begin with, it is not totally apparent if a PBH can be formed from an isocurvature 
perturbation of a massless scalar field. To formulate the problem in a consistent way,
we have to set the appropriate initial condition for a primordially
 isocurvature perturbation. 
For this purpose, we employ the spatial gradient expansion,
which is valid for perturbations on scales much larger than the Hubble horizon scale.
Thus we first perform the long-wavelength expansion of the field equations
by taking the existence of the massless scalar field into account.
Due to the isocurvature nature of perturbations,
the zeroth order solution for the metric should be given by 
the standard homogeneous and isotropic radiation-dominated universe.
Thus we can easily identify the leading order isocurvature perturbation and obtain 
the general solution for the isocurvature growing mode up through the 
next leading order terms in gradient expansion, which arise from the spatial 
gradients of the scalar field. 
These next leading order terms are the ones that gradually induce 
non-trivial perturbations in the geometry, and may eventually lead to PBH formation.
In fact, we explicitly show that this isocurvature growing mode
can result in the formation of a PBH if the initial amplitude $\mu$
is sufficiently large.

This paper is organized as follows. 
We write down all relevant equations in the form of the 
cosmological 3+1 decomposition in Sec.~\ref{sec:eom31} 
following Refs.~\cite{Shibata:1999zs,Harada:2015yda}. 
Then, performing the gradient expansion, we derive the growing mode solution 
for a massless scalar field isocurvature perturbation in Sec.~\ref{sec:graex}.  
The spherically symmetric numerical simulations are described in Sec.~\ref{sec:numeres}. 
The numerical code, the initial data, the time evolution of dynamical quantities, 
the evolution of the PBH mass after the formation, and 
the scaling of the black hole mass near the threshold are discussed in 
Secs.~\ref{sec:numecod} -- \ref{sec:critical}, respectively. 
Sec.~\ref{sec:sumdis} is devoted to a summary and discussion. 

Throughout this paper, we use the geometrized units in which both 
the speed of light and Newton's gravitational constant are set to unity, $c=G=1$.

%%%%%%%%%%%%%%%%%%%%%%%%%%%%%%%%%%%%%%%%%%%%%%%%%%%%%%%%%%%%%%%%
\section{Long-wavelength approximation for a perfect fluid-massless scalar system}
%%%%%%%%%%%%%%%%%%%%%%%%%%%%%%%%%%%%%%%%%%%%%%%%%%%%%%%%%%%%%%%%

\subsection{Equations of motion}
\label{sec:eom31}
Let us write down the cosmological 3+1 form of the equations of motion 
given in Ref.~\cite{Harada:2015yda} adding a massless scalar field. 
Since there is no direct coupling between the scalar field and the fluid, 
the equations of motion can be obtained by simply
adding the contribution of the massless scalar field into the stress-energy tensor 
and additionally considering the dynamical equations for the massless scalar field.  
In the following, we recapitulate the Einstein equations
and the conservation and Euler equations from Ref.~\cite{Harada:2015yda}
(see also \cite{Shibata:1999zs}).
We use the notation in which the Greek indices run over 0 to 3, and 
the Latin indices over 1 to 3. 
The reference spatial metric associated with the coordinates $x^i$ will be 
expressed as $\eta_{ij}$, where $x^i$ are not necessarily Cartesian.
The Cartesian coordinates are denoted by $X^i=(X,Y,Z)$. 

The Hamiltonian and momentum constraints are given as follows: 
\begin{eqnarray}
&& \tilde{\triangle}\psi=\frac{\tilde{\cal R}_{k}^{k}}{8}\psi-2\pi
  \psi^{5}a^{2}E
-\frac{\psi^{5}a^{2}}{8}\left(\tilde{A}_{ij}\tilde{A}^{ij}-\frac{2}{3}K^{2}\right), \label{eq:2.37}\\
&&
 \tilde{\cal D}^{j}(\psi^{6}\tilde{A}_{ij})-\frac{2}{3}\psi^{6}\tilde{\cal D}_{i}K=8\pi
 J_{i}\psi^{6}, 
\label{eq:2.38}
\end{eqnarray}
where $\tilde \triangle$ is the Laplacian with respect to the conformal metric $\tilde \gamma_{ij}$ which is related to 
the spatial metric $\gamma_{ij}$ as $\gamma_{ij}=\psi^4a^2\tilde \gamma_{ij}$ with $\det(\tilde \gamma)=\eta:=\det(\eta)$ and 
$a$ being the scale factor in the reference universe. 
$\tilde {\mathcal R}_{ij}$ is the Ricci tensor with respect to the conformal metric $\tilde \gamma_{ij}$. 
$E$ and $J_i$ are the energy density and momentum density for the Eulerian observer defined by 
$E:=n_\mu n_\nu T^{\mu\nu}$ and $J_i:=-\gamma_{i\mu}n_\nu T^{\mu\nu}$ with $n_\mu$ and $T^{\mu\nu}$ being 
the unit vector normal to the time slice and total stress-energy tensor, respectively. 
$\tilde A_{ij}$ is related to the traceless part $A_{ij}$ of the extrinsic curvature $K_{ij}$ as $A_{ij}=\psi^4a^2\tilde A_{ij}$. 
We raise the Latin lowercase indices $i,j,k,...$ of tilded quantities 
by $\tilde \gamma^{ij}$, and $\tilde {\mathcal D}_i$ is 
the covariant derivative with respect to $\tilde \gamma_{ij}$. 

The evolution equations for $\psi$, $\tilde \gamma_{ij}$, $K=\gamma^{ij}K_{ij}$ and $\tilde A_{ij}$ are given as
\begin{eqnarray}
(\partial_{t}-{\cal L}_{\beta})\psi &=&
 -\frac{\dot{a}}{2a}\psi+\frac{\psi}{6}(-\alpha K+\bar{\cal D}_{k}\beta^{k}), 
\label{eq:2.40} 
\\
 (\partial_{t}-{\cal L}_{\beta})\tilde{\gamma}_{ij}&=&-2\alpha
  \tilde{A}_{ij}-\frac{2}{3}\tilde{\gamma}_{ij}\bar{\cal D}_{k}\beta^{k}.
\label{eq:2.44}
\\
(\partial_{t}-{\cal L}_{\beta})K &=&
 \alpha\left(\tilde{A}_{ij}\tilde{A}^{ij}+\frac{1}{3}K^{2}\right)-D_{k}D^{k}\alpha
 +4\pi\alpha (E+S_{k}^{k}), 
\label{eq:2.41}
\\
 (\partial_{t}-{\cal L}_{\beta})\tilde{A}_{ij}&=&\frac{1}{a^{2}\psi^{4}}
\left[\alpha\left({\cal R}_{ij}-\frac{\gamma_{ij}}{3}{\cal R}\right)
-\left(D_{i}D_{j}\alpha-\frac{\gamma_{ij}}{3}D_{k}D^{k}\alpha\right)\right]
\nonumber \\
&& +\alpha(K\tilde{A}_{ij}-2\tilde{A}_{ik}\tilde{A}_{j}^{k})
-\frac{2}{3}(\bar{\cal D}_{k}\beta^{k})\tilde{A}_{ij}
-\frac{8\pi\alpha}{a^{2}\psi^{4}}\left(S_{ij}-\frac{\gamma_{ij}}{3}S_{k}^{k}\right), 
\label{eq:2.39}
\end{eqnarray}
where $\alpha$ is the lapse function, 
$\mathcal L_\beta$ is the Lie derivative associated with the shift vector $\beta^i$, and 
$D_i$ is the covariant derivative with respect to $\gamma_{ij}$. 

The conservation of the stress-energy tensor for a perfect fluid can be rewritten as 
follows:
\begin{eqnarray}
&&\left[\psi^{6}a^{3}\left\{(\rho+p)\Gamma^{2}-p\right\}\right]_{,t}
+\frac{1}{\sqrt{\eta}}\left[\sqrt{\eta}
\psi^{6}a^{3}\left\{(\rho+p)\Gamma^{2}-p\right\}v^{l}\right]_{,l}
\nonumber \\
&&=-\frac{1}{\sqrt{\eta}}\left[\sqrt{\eta}\psi^{6}a^{3}p(v^{l}+\beta^{l})\right]_{,l}
+\alpha\psi^{6}a^{3}pK
-\alpha^{-1}\alpha_{,l}\psi^{6}a^{3}\Gamma^{2}(\rho+p)(v^{l}+\beta^{l})
\nonumber \\
&&
 +\alpha^{-1}\psi^{10}a^{5}\Gamma^{2}(\rho+p)(v^{l}+\beta^{l})(v^{m}+\beta^{m})\left(\tilde{A}_{lm}+\frac{\tilde{\gamma}_{lm}}{3}K\right), 
\label{eq:2.45}
\\
&&(\Gamma\psi^{6}a^{3}(\rho+p)u_{j})_{,t}+\frac{1}{\sqrt{\eta}}(\sqrt{\eta}
\Gamma\psi^{6}a^{3}(\rho+p)v^{k}u_{j})_{,k} \nonumber \\
&&=-\alpha\psi^{6}a^{3}p_{,j}+\Gamma\psi^{6}a^{3}(\rho+p)
\left(-\Gamma\alpha_{,j}+u_{k}\beta^{k}_{,j}-\frac{u_{k}u_{l}}{2u^{t}}\gamma^{kl}_{,j}\right),
\label{eq:2.46}
\\
&&(\Gamma\psi^{6}a^{3}n)_{,t}+\frac{1}{\sqrt{\eta}}
(\sqrt{\eta}\Gamma\psi^{6}a^{3}nv^{k})_{,k}=0, 
\label{eq:2.47}
\end{eqnarray}
where $\rho$, $p$ and $u^\mu$ are the energy density, pressure and the four-velocity,
respectively, of the fluid, by which the fluid stress-energy tensor is given
as $T^{\rm f}_{\mu\nu}=(\rho+p)u_\mu u_\nu+p g_{\mu\nu}$, 
and $v^i$ and $\Gamma$ 
are defined by $v^i=u^i/u^t$ and $\Gamma=\alpha u^t$, respectively.
\footnote{$\Gamma$ is denoted by $w$ in Ref.~\cite{Harada:2015yda}}.

The field equations for a massless scalar field $\phi$ are written as 
\begin{eqnarray}
(\del_t-\beta^i\del_i)\phi&=&-\alpha \varpi, \\
(\del_t-\beta^i\del_i)\varpi&=&-\alpha \triangle \phi -\gamma^{\mu\nu}\del_\mu\alpha \del_\nu\phi+\alpha K \varpi, 
\end{eqnarray}
where $\triangle$ is the Laplacian with respect to the spatial metric $\gamma_{ij}$. 
As the stress-energy tensor of $\phi$ is given by 
$T^{\rm sc}_{\mu\nu}=\nabla_\mu \phi \nabla_\nu\phi-\frac{1}{2}g_{\mu\nu}\nabla^\lambda\phi\nabla_\lambda\phi$, 
its contribution to $E$, $J_i$ and the stress-tensor $S_{ij}$ are given by 
\begin{eqnarray}
E^{\rm sc}&=&n^\mu n^\nu T^{\rm sc}_{\mu\nu}=\frac{1}{2}\varpi^2+\frac{1}{2}\psi^{-4}a^{-2}\tilde \gamma^{ij}\del_i\phi\del_j\phi, 
\label{eq:Esc}\\
J^{\rm sc}_i&=&-\gamma_i^{~\mu}n^\nu T^{\rm sc}_{\mu\nu}=\varpi \del_i \phi, \\
S^{\rm sc}_{ij}&=&=\gamma_i^{~\mu}\gamma_j^{~\nu}T^{\rm sc}_{\mu\nu}=\del_i\phi\del_j\phi
%-\psi^4a^2\tilde\gamma_{ij}V
-\frac{1}{2}\tilde\gamma_{ij}\tilde\gamma^{kl}\del_k\phi\del_l\phi
+\frac{1}{2}\psi^4a^2\tilde \gamma_{ij}\varpi^2. 
\end{eqnarray}

\subsection{Gradient expansion}
\label{sec:graex}
We apply the spatial gradient expansion to the equations of motion by introducing 
the fictitious expansion parameter $\epsilon$ associated with the spatial derivative. 
To be precise, we put the parameter $\epsilon$ in front of the spatial partial derivative,
e.g., $\del_i\rightarrow \epsilon \del_i$. 
For a typical wave number $k$ associated with an inhomogeneity, 
$\epsilon\sim k/(aH_{\rm b})$, where $H_{\rm b}$ is the background Hubble parameter. 
In this paper, we assume that the massless scalar field does not 
contribute to the background metric of the gradient expansion.
That is, the only matter component in the background FLRW universe is the fluid,
which we leave unspecified, though 
we take it to be the radiation in the numerical simulation. 
Since the massless scalar can contribute to the stress-energy tensor only 
through the derivative terms, 
we may assume the massless scalar field to have the form,
\begin{equation}
\phi=\Upsilon(\bm x)+\lambda(t,\bm x), 
\end{equation}
where $\Upsilon$ is assumed to be $\mathcal O(1)$ 
and $\lambda(t,\bm x)=\mathcal O(\epsilon)$. 
This is the most general form of $\phi$ which makes no leading order 
contribution to the stress energy tensor, Eq.~\eqref{eq:Esc}. 
Then, expanding the scalar field equations, we obtain 
\begin{eqnarray}
&&\del_t\lambda=-\varpi+\mathcal O(\epsilon^2), 
\label{eq:ms1}\\
&&\del_t\varpi=K_{\rm b}\varpi +\mathcal O(\epsilon^2). 
\label{eq:ms2}
\end{eqnarray}
Using the fact that $K_{\rm b}=-3\dot a/a$, we obtain the general solution to Eqs.~\eqref{eq:ms1} and \eqref{eq:ms2} as follows: 
\begin{equation}
\lambda=\lambda_1(\bm x)+\lambda_2(\bm x) \int a^{-3} \dd t+\mathcal O(\epsilon^2). 
\label{eq:phi}
\end{equation}
The second term is the decaying mode, and we drop it. 
Since the first term can be absorbed into the leading term $\Upsilon(\bm x)$, 
we conclude 
\begin{equation}
\lambda(t,\bm x)=\mathcal O(\epsilon^2). 
\end{equation}

With the above result for the massless scalar field, we find
the order counting for the other variables are unchanged 
from Ref.~\cite{Harada:2015yda}. 
To summarize, the expansion orders of the variables are given as follows:
\begin{eqnarray}
\psi&=&\Psi(\bm x)(1+\xi(t,\bm x))+\mathcal O(\epsilon^3), \\
\tilde \gamma_{ij}&=&\eta_{ij}+h_{ij}+\mathcal O(\epsilon^3), \\
K&=&K_{\rm b}(1+\kappa)+\mathcal O(\epsilon^3), \\
\alpha&=&1+\chi+\mathcal O(\epsilon^3), \\
\beta^i&=&\mathcal O(\epsilon), \\
\rho&=&\rho_{\rm b}(1+\delta)+\mathcal O(\epsilon^3), \\
v^i&=&\mathcal O(\epsilon), \\
\phi&=&\Upsilon(\bm x)+\lambda(t,\bm x)+\mathcal O(\epsilon^3), \\
\varpi&=&\mathcal O(\epsilon^2), 
\end{eqnarray}
where $K_{\rm b}=-3H_{\rm b}$, $\rho_{\rm b}=3H_{\rm b}^2/(8\pi)$, and 
$\Psi=\mathcal O(1)$. The perturbation variables 
$\xi$, $h_{ij}$, $\kappa$, $\chi$ and $\delta$ are of $\mathcal O(\epsilon^2)$. 
We also find $v^i+\beta^i=\mathcal O(\epsilon^3)$ 
(see Ref.~\cite{Harada:2015yda} for details). 

Hereafter, we fix the gauge as the uniform-mean-curvature slicing~($\kappa=0$) 
and the normal coordinates~($\beta^i=0$), 
so that $v^i=\mathcal O(\epsilon^3)$. 
We focus on the perfect fluid with the equation of state $p=w\rho$ with constant $w$.
Since we are interested in primordially isocurvature perturbations,
 we may set $\Psi=1$, implying that there is no curvature perturbation initially.
Then, we obtain the following set of equations:
\begin{eqnarray}
&&\dot \delta+6\dot \xi+3H_{\rm b}w\chi=\mathcal O(\epsilon^4), 
\label{eq:4.12ms}\\
&&\frac{1}{1+w}\dot \delta+6\dot \xi=\mathcal O(\epsilon^4), 
\label{eq:4.13ms}\\
&&\del_t[a^3(1+w)\rho_{\rm b}u_j]=-a^3\rho_{\rm b}[w\del_j\delta+(1+w)\del_j \chi]+\mathcal O(\epsilon^5), 
\label{eq:4.15ms}
\\
&&6\dot \xi-3H_{\rm b}\chi=\mathcal O(\epsilon^4), 
\label{eq:4.16ms}\\
&&\rho_{\rm b}\delta+\frac{1}{2}a^{-2}(\mathcal D\Upsilon)^2=\mathcal O(\epsilon^4), 
\label{eq:4.17ms}\\
&&3(1+w)\chi+(1+3w)\delta =\mathcal O(\epsilon^4), 
\label{eq:4.18ms}\\
&&\del_th_{ij}=-2\tilde A_{ij}+\mathcal O(\epsilon^4), 
\label{eq:4.19ms}\\
&&\del_t \tilde A_{ij}+\frac{3\dot a}{a}\tilde A_{ij}=-\frac{8\pi}{a^2}\left(\mathcal D_i\Upsilon\mathcal D_j\Upsilon-\frac{1}{3}\eta_{ij}(\mathcal D\Upsilon)^2\right)+\mathcal O(\epsilon^4), 
\label{eq:4.20ms}\\
&&\bar {\mathcal D}_i(\Psi\tilde A^i_j)=8\pi(1+w)\rho_{\rm b}u_j+8\pi\varpi\mathcal D_i\Upsilon+\mathcal O(\epsilon^5), 
\label{eq:4.21ms}\\
&&\del_t\lambda=-\varpi+\mathcal O(\epsilon^3), 
\label{eq:s1ms}\\
&&\del_t\varpi=-a^{-2}\bar\triangle\Upsilon-3H_{\rm b}\varpi, 
\label{eq:s2ms}
\end{eqnarray}
where $(\mathcal D\Upsilon)^2=\eta^{ij}\mathcal D_i\Upsilon \mathcal D_j\Upsilon$ and $\bar \triangle=\eta^{ij}\mathcal D_i\mathcal D_j$. 

Let us solve the above set of equations.
The initial condition is given by the scalar
field configuration $\phi=\Upsilon(\bm x)$.
From Eqs.~\eqref{eq:4.17ms} and \eqref{eq:4.18ms}, we find 
\begin{eqnarray}
\delta&=&-\frac{4\pi}{3}\left(\frac{1}{aH_{\rm b}}\right)^2(\mathcal D\Upsilon)^2,
\\
\chi&=&-\frac{1+3w}{3(1+w)}\delta. 
\end{eqnarray}
From Eq.~\eqref{eq:4.16ms}, we obtain 
\begin{equation}
\dot \xi=\frac{2\pi}{9}\frac{1+3w}{1+w}\frac{1}{a}\left(\frac{1}{aH_{\rm b}}\right)(\mathcal D\Upsilon)^2.
\end{equation}
Integrating this equation and ignoring the constant mode, 
which corresponds to a non-vanishing curvature perturbation, we obtain  
\begin{equation}
\xi=\frac{2\pi}{9}\frac{1}{1+w}\left(\frac{1}{aH_{\rm b}}\right)^2(\mathcal D\Upsilon)^2, 
\end{equation}
where we have used the relation:
$$\frac{\dd}{\dd t}\left(\frac{1}{aH_{\rm b}}\right)^2=\frac{1+3w}{a^2H_{\rm b}}. $$
One can directly check that the above expressions solve
 Eqs.~\eqref{eq:4.12ms} and \eqref{eq:4.13ms}. 
Thus although we have assumed $\Psi=1$, $\xi$ at $\mathcal O(\epsilon^2)$
 does not vanish in general, and may lead to 
a significant metric perturbation around the horizon entry.

Integrating Eqs.~\eqref{eq:4.15ms}, \eqref{eq:4.21ms}, \eqref{eq:4.19ms} 
and \eqref{eq:s2ms}, and dropping the decaying modes, 
we obtain 
\begin{eqnarray}
u_i&=&-\frac{8\pi}{9}\frac{a}{(1+w)(5+3w)}\left(\frac{1}{aH_{\rm b}}\right)^3\mathcal D_i(\mathcal D\Upsilon)^2, 
\label{eq:ui}
\\
\tilde A_{ij}&=&-\frac{16\pi}{5+3w}\frac{1}{a}\frac{1}{aH_{\rm b}}\left[\mathcal D_i\Upsilon\mathcal D_j\Upsilon-\frac{1}{3}\eta_{ij}(\mathcal D\Upsilon)^2\right], 
\label{eq:Aij}
\\
h_{ij}&=&\frac{32\pi}{(5+3w)(1+3w)}\left(\frac{1}{aH_{\rm b}}\right)^2\left[\mathcal D_i\Upsilon\mathcal D_j\Upsilon-\frac{1}{3}\eta_{ij}(\mathcal D\Upsilon)^2\right], 
\label{eq:hij}
\\
\varpi&=&-\frac{2}{5+3w}\frac{1}{a}\frac{1}{aH_{\rm b}}\bar\triangle \Upsilon. 
\label{eq:Pi}
\end{eqnarray}
By using Eqs.~\eqref{eq:ui}, \eqref{eq:Aij} and \eqref{eq:Pi}, 
one can show that the momentum constraint Eq.~\eqref{eq:4.21ms} is satisfied. 
Finally, from Eq.~\eqref{eq:s1ms}, $\lambda$ is given by 
\begin{equation}
\lambda=\frac{2}{(5+3w)(1+3w)}\left(\frac{1}{aH_{\rm b}}\right)^2\bar\triangle \Upsilon. 
\label{eq:lam}
\end{equation}

As is explicitly shown in this subsection, 
once the functional form of $\Upsilon(\bm x)$ is specified, 
we can solve the equations of motion at each order in the gradient expansion. 
To summarize, the solution with $\Psi(\bm x)=1$ but non-trivial $\Upsilon(\bm x)$
is the isocurvature mode, which is of our interest, while the one with 
$\Upsilon(\bm x)={\rm const.}$ but non-trivial $\Psi(\bm x)$ 
corresponds to the standard curvature perturbation. 
To be more precise, general growing mode solutions are generated by 
the set of the functions $\Psi(\bm x)$ and $\Upsilon(\bm x)$ whose spatial variations are initially in the super-horizon regime. 
As $\Upsilon(\bm x)$ does not contribute to the spacetime curvature up until $O(\epsilon^4)$, while $\Psi(\bm x)$ contributes at $O(\epsilon^2)$, 
we focus on the perturbations independent of those generated from $\Psi(\bm x)$, which is the definition of the isocurvature mode in this paper. 

Before ending this subsection, it is worth defining and evaluating the compaction function of our isocurvature initial condition. 
Although the distinction of the curvature perturbation component from the isocurvature one at the horizon entry is difficult due to the nonlinear mixing, 
it is still possible to consider the compaction function for the isocurvature perturbation on superhorizon scales 
when the curvature component is negligible. 
Let us define the compaction function for the isocurvature perturbation
by $\mathcal C(R)\equiv 2\delta M(R)/R$ where $\delta M(R)$ is the energy carried by the scalar field within the radius $R$. 
  That is, since we have $E^{\rm SC}=\frac{1}{2}a^{-2}(\del_r\Upsilon)^2+\mathcal O(\epsilon^4)$, we define the compaction function by 
  \begin{equation}
    \mathcal C(R)=\frac{4\pi}{R}\int^{R}_0 (\del_r\Upsilon)^2 r^2\dd r. 
  \end{equation}
For our initial condition, 
\begin{eqnarray}
  \mathcal C(R)
=\frac{4\pi\mu^2}{9R}k^4\int_0^R\exp\left[-\frac{k^2r^2}{3}\right]r^4dr\,.
\end{eqnarray}
The maximum of $\mathcal C(R)$ is at 
$kR\sim 3.5$
with the value $\mathcal C\sim 3.5\mu^2$. 
Thus if we apply a similar criterion used for an adiabatic perturbation and assume $\mathcal C_{\rm critical}\sim1$ for PBH formation, 
the corresponding critical value of $\mu$ would be $\mu\sim 0.53$. 
As we shall see in the next subsection, 
this value is fairly close to the actual critical value. 
This indicates that the compaction function criterion is also useful for initially isocurvature perturbations.

%%%%%%%%%%%%%%%%%%%%%%%%%%%%%%%%%%%%%%%%%%%%%%%%%%%%%%%%%%%%%%%%
\section{Numerical simulation}
\label{sec:numeres}
%%%%%%%%%%%%%%%%%%%%%%%%%%%%%%%%%%%%%%%%%%%%%%%%%%%%%%%%%%%%%%%%

%%%%%%%%%%%%%%%%%%%%%%%%%%%%%%%%%%%%%%%%%%%%%%%%%%%%%%%%%%%%%%%%
\subsection{Numerical code}
\label{sec:numecod}
%%%%%%%%%%%%%%%%%%%%%%%%%%%%%%%%%%%%%%%%%%%%%%%%%%%%%%%%%%%%%%%%
Numerical simulations are performed by using newly developed numerical code based on 
the 3+1 dimensional simulation code used in Refs.~\cite{Yoo:2018pda,Yoo:2020lmg}. 
The 3+1 dimensional numerical code is 
modified by implementing the CARTOON method~\cite{Alcubierre:1999ab},
so that the new code is specialized for spherically symmetric systems. 
That is, we solve the evolution equations only on the $Z$-axis and create the necessary data around it for the time evolution 
based on the spherical symmetry. 
We also set the background radiation-dominated, $w=1/3$.
For this purpose, we need to evaluate the physical quantities at off-grid points on the
$Z$-axis with interpolation. 
We implemented the 3rd order spline interpolation to realize the stable time evolution. 
As in Refs.~\cite{Yoo:2018pda,Yoo:2020lmg}, 
we use the scale-up coordinates with the parameter $\eta$, 
where the ratio of the unit coordinate interval in the Cartesian 
coordinates near the center is finer by a factor $1+2\eta$ in comparison 
to that near the boundary~(see Ref.~\cite{Yoo:2018pda} for details). 
The value of $\eta$ is set $\eta=40$ except for the simulation shown 
in Sec.~\ref{sec:critical}, where we set $\eta=20$ 
with fixed mesh refinement. 
When a black hole region is observed, 
we excise a part of the inner region in the subsequent time evolution.

%%%%%%%%%%%%%%%%%%%%%%%%%%%%%%%%%%%%%%%%%%%%%%%%%%%%%%%%%%%%%%%%
\subsection{Initial setting}
\label{sec:initial}
%%%%%%%%%%%%%%%%%%%%%%%%%%%%%%%%%%%%%%%%%%%%%%%%%%%%%%%%%%%%%%%%
As is shown in Sec.~\ref{sec:graex}, 
once we specify the functional form of $\Upsilon$, we obtain the long-wavelength solution 
for an isocurvature perturbation. 
In this paper, we consider the following specific spherically 
symmetric profile of $\Upsilon$:
\begin{equation}
\Upsilon(R)=\mu\exp\left(-\frac{1}{6}k^2R^2\right)W(R), 
\label{eq:scg}
\end{equation}
where $R^2=X^2+Y^2+Z^2$ with $X$, $Y$ and $Z$ being the Cartesian coordinates for $\eta_{ij}$, 
and the function $W(R)$ is introduced to make $\phi$ smoothly vanish at the boundary. 
The specific form of $W(R)$ is given by~\cite{Yoo:2018pda} 
\begin{equation}
  W(R)=
  \left\{
  \begin{array}{ll}
  1&{\rm for}~0\leq R \leq R_{\rm W} \\
  1-\frac{\left(\left(R_{\rm W}-L\right)^6-\left(L-R\right)^6\right)^6}{\left(R_{\rm W}-L\right)^{36}}&{\rm for}~R_{\rm W} \leq R \leq L 
  \\
  0&{\rm for}~L \leq R 
  \end{array}\right.,
  \end{equation}  
  where $L$ is the Cartesian radius at the numerical outer boundary and $R_W$ is set to $4L/5$. 
The amplitude of the inhomogeneity is characterized by the parameter $\mu$, 
and $1/k$ gives the scale of the inhomogeneity because 
$\triangle \Upsilon|_{r=0}=-k^2\Upsilon$. 
It may be noted, however, if we simply identify 
the Gaussian profile with the Gaussian window function adopted in the PBH formation
from the curvature perturbation, the corresponding comoving wavenumber would be
$k_*^2=k^2/3$ because the standard Gaussian window function associated with the wave number $k_*$ is defined as $\exp(-k^2/(2k_*^2))$. 

We adopt the asymptotically FLRW boundary condition used in Ref.~\cite{Shibata:1999zs},
but the constraints are significantly violated on the boundary through 
the time evolution%
\footnote{The constraint violation at the boundary starts to occur when the effects of the inhomogeneity reach the boundary, and  
the origin of the constraint violation is due to unphysical reflected modes. 
As far as we know, for the asymptotically FLRW boundary, 
no established procedure to reduce the unphysical reflected modes has been reported. 
}. 
Therefore we terminate the time evolution before the boundary effects 
influence the formation and evolution of a PBH. 
At the initial time, the Cartesian radius $L$ at the numerical outer boundary is set to 1 which corresponds to the 
normalization of the scale factor.  
We fix the ratio between the wave number $k$ and the initial Hubble parameter $H_{\rm i}$ as $H_{\rm i}/k=5$. 
The initial value of the background Hubble parameter 
is set to $50/L$ or $100/L$ depending on the resolution requirement. 
Namely,we adopt $50/L$ for simulating the PBH formation process, 
while we adopt $100/L$ for evolving the dynamics after the PBH formation.
A smaller value of $H_{\rm i}$ gives a finer resolution at the center, 
but the effects of the boundary reach the central region earlier. 
We define the horizon entry time by $t_{\rm H}$ by $aH_{\rm b}|_{t=t_{\rm H}}=k$, 
and the horizon mass by
 $M_{\rm H}=1/(2H_{\rm b}(t_{\rm H}))=t_{\rm H}$.\footnote{
As noted in the above,
if we use $k_*$ to define the horizon entry in place of $k$, 
the corresponding time, and hence the horizon mass,
 becomes 3 times larger, $t_*=3t_{\rm H}$.}

%%%%%%%%%%%%%%%%%%%%%%%%%%%%%%%%%%%%%%%%%%%%%%%%%%%%%%%%%%%%%%%%
\subsection{Time evolution of two typical cases}
\label{sec:timeevo}
\begin{figure}[htbp]
\begin{center}
\includegraphics[scale=0.92]{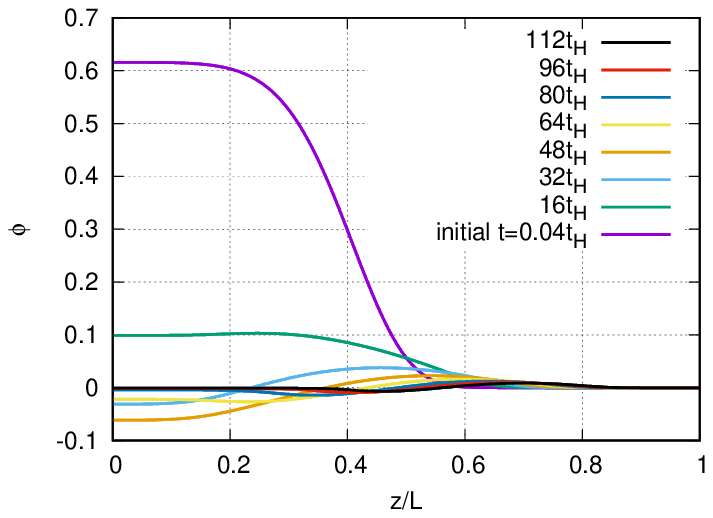}
\includegraphics[scale=0.92]{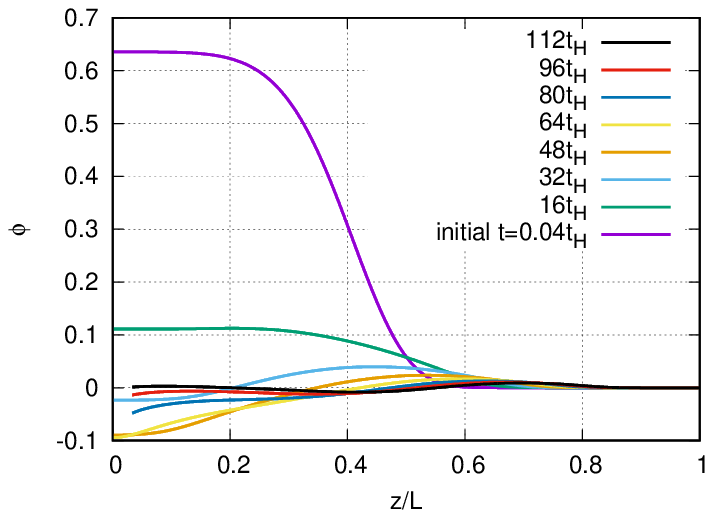}
\\
\includegraphics[scale=0.92]{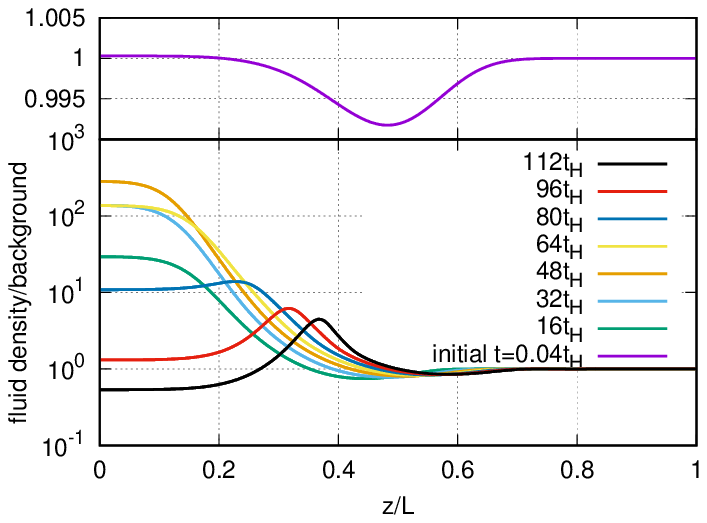}
\includegraphics[scale=0.92]{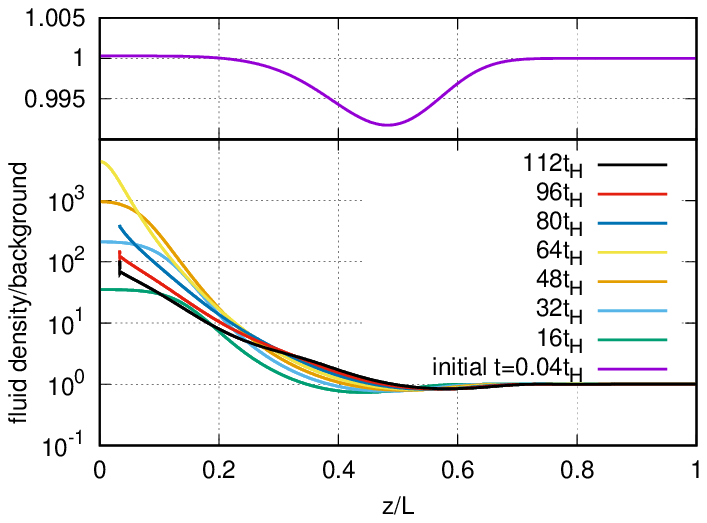}
\\
\includegraphics[scale=0.92]{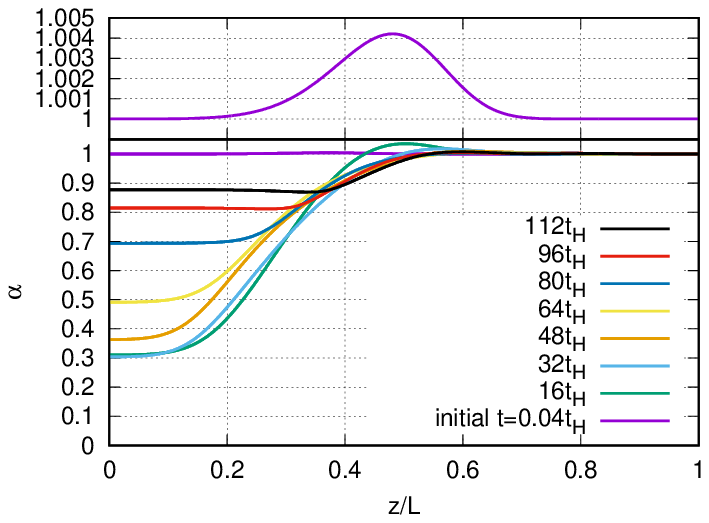}
\includegraphics[scale=0.92]{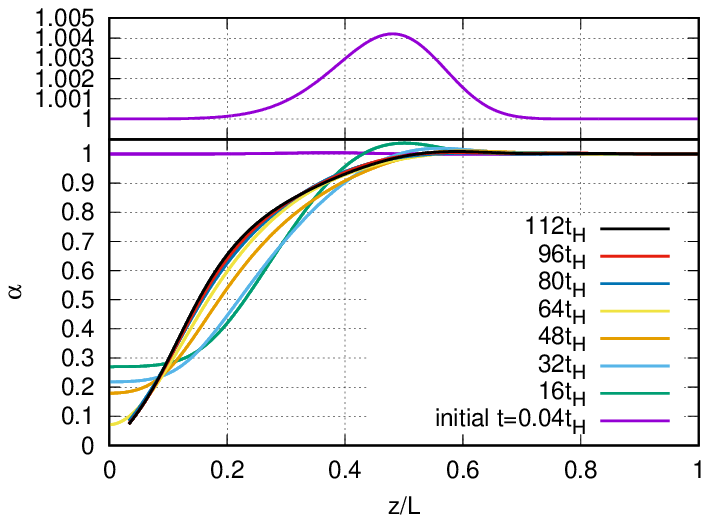}
\caption{\baselineskip5mm
Snapshots of the lapse function, scalar field, ratio of the fluid density to the background 
are shown for $\mu=0.62$ (left) and $\mu=0.64$ (right). 
In the lower two rows, the initial configuration is shown in a different scale. 
}
\label{fig:evo_figs}
\end{center}
\end{figure}

\begin{figure}[htbp]
\begin{center}
\includegraphics[scale=1]{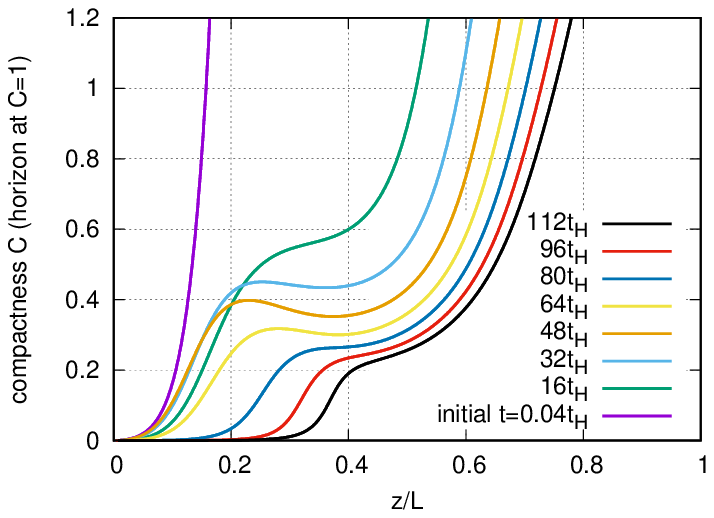}
\includegraphics[scale=1]{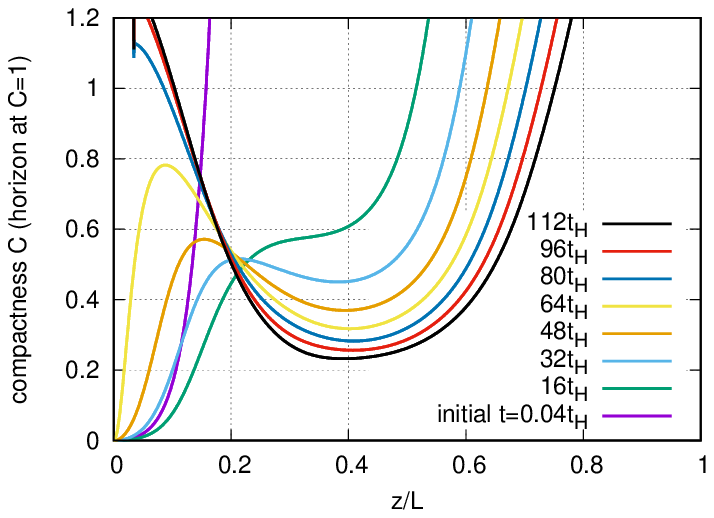}
\caption{\baselineskip5mm
Snapshots of the compactness $C$(Misner-Sharp mass/area radius) 
are shown for $\mu=0.62$ (left) and $\mu=0.64$ (right). 
For the $\mu=0.64$ case, the black hole horizon is formed at around $t=74 t_{\rm H}$. 
}
\label{fig:evo_comp}
\end{center}
\end{figure}

%%%%%%%%%%%%%%%%%%%%%%%%%%%%%%%%%%%%%%%%%%%%%%%%%%%%%%%%%%%%%%%%
First, let us show two typical cases $\mu=0.62$ and $\mu=0.64$ in 
Figs.~\ref{fig:evo_figs}, \ref{fig:evo_comp} and \ref{fig:const}. 
Snapshots of the scalar field, the ratio of the fluid density to the background value,
and the lapse function are shown in Fig.~\ref{fig:evo_figs}. 
It should be noted that the horizontal axis is the scale-up coordinate $z$, and 
the initial profile of the scalar field is not Gaussian in this coordinate 
although it is Gaussian in the coordinate $R$ 
as is given in Eq.~\eqref{eq:scg}. 
The scalar field initially contributes to the energy density mainly through 
its spatial gradients, and the energy excess due to the gradient energy 
is compensated by the under density of the fluid because of
the isocurvature condition, $\Psi=1$ (see Eq.~\eqref{eq:4.17ms}). 
The scalar field decays after the horizon entry in both cases. 
%%%%%%%%%%%%%%%%%%%%%%%%%%%<<start figure>>%%%%%%%%%%%%%%%%%%%%%%%%%%
%%%%%%%%%%%%%%%%%%%%%%%%%%%%<<end figure>>%%%%%%%%%%%%%%%%%%%%%%%%%%%

In Fig.~\ref{fig:evo_comp}, the compactness $C$~(Misner-Sharp mass $M_{\rm MS}$ divided 
by the half of the area radius) is plotted as a function of $z$ at several moments of time. 
For the $\mu=0.64$ case, the fluid energy density forms a sharp peak
at the center~(see right-middle panel in Fig.~\ref{fig:evo_figs}) and 
a black hole apparent horizon~($C=1$ and $\del_r C<0$ in Fig.~\ref{fig:evo_comp}) appears 
as is explicitly shown in Fig.~\ref{fig:evo_comp}. 
After the black hole formation, we excise a central region inside the horizon. 
For the $\mu=0.62$ case, no black hole formation is observed, and we 
find that the density wave of the fluid propagates outwards. 
The lapse function for the $\mu=0.62$ case bounces back to the value close to 1. 
In Fig.~\ref{fig:evo_comp}, we also find that the cosmological 
horizon~($C=1$ and $\del_r C>0$ in Fig.~\ref{fig:evo_comp}) 
always exists and expands outwards. 
%%%%%%%%%%%%%%%%%%%%%%%%%%%<<start figure>>%%%%%%%%%%%%%%%%%%%%%%%%%%
%%%%%%%%%%%%%%%%%%%%%%%%%%%%<<end figure>>%%%%%%%%%%%%%%%%%%%%%%%%%%%

\begin{figure}[htbp]
\begin{center}
\includegraphics[scale=1]{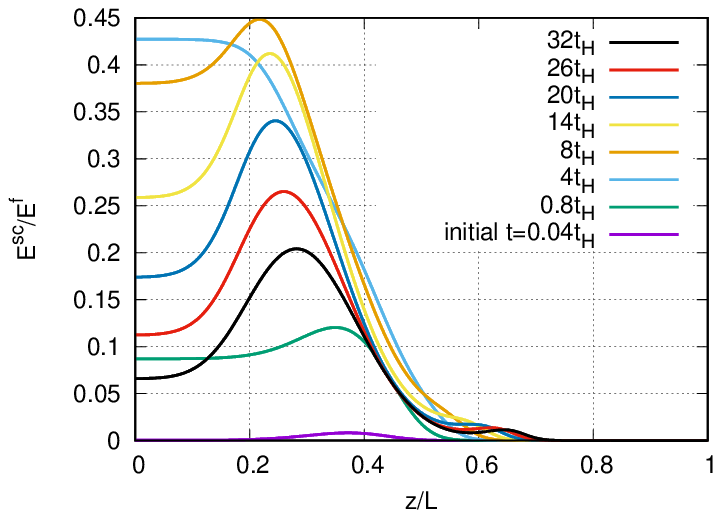}
\includegraphics[scale=1]{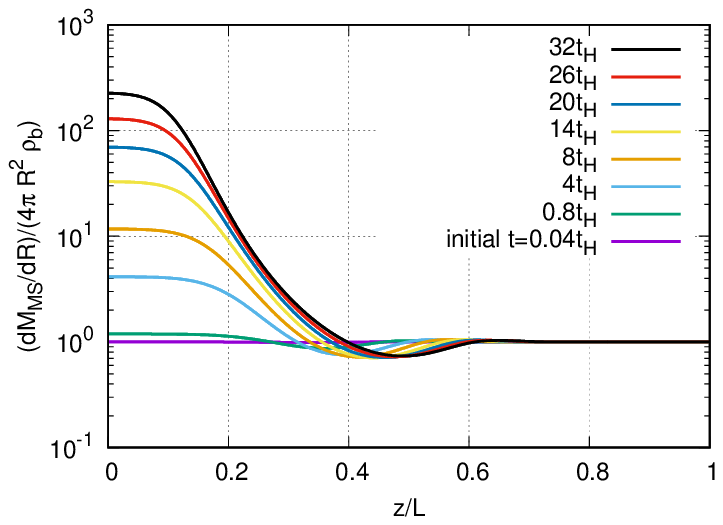}\caption{\baselineskip5mm
Snapshots of the ratio of the scalar field energy density to 
the fluid energy density $E^{\rm sc}/E^{\rm f}$ (left) and the ratio of 
the total energy density 
to the background energy density $\dd M_{\rm MS}/\dd R_{\rm A} / (4\pi R_{\rm A}^2\rho_{\rm b})$,
for $\mu=0.64$.
}
\label{fig:eneden}
\end{center}
\end{figure}

\begin{figure}[htbp]
\begin{center}
\includegraphics[scale=1]{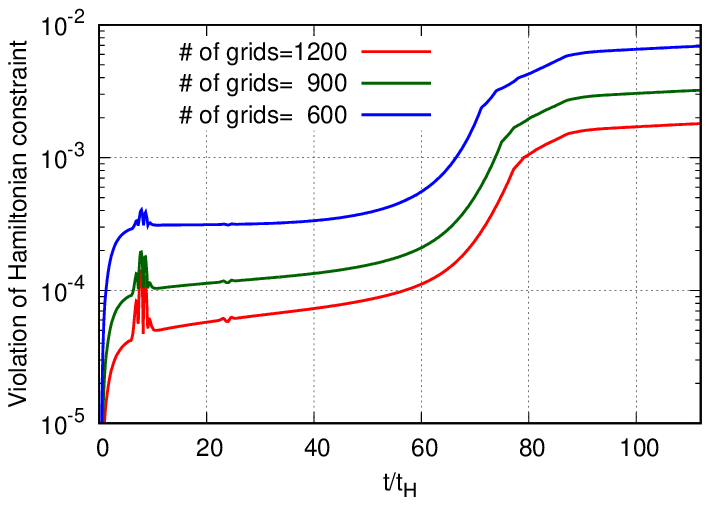}
\includegraphics[scale=1]{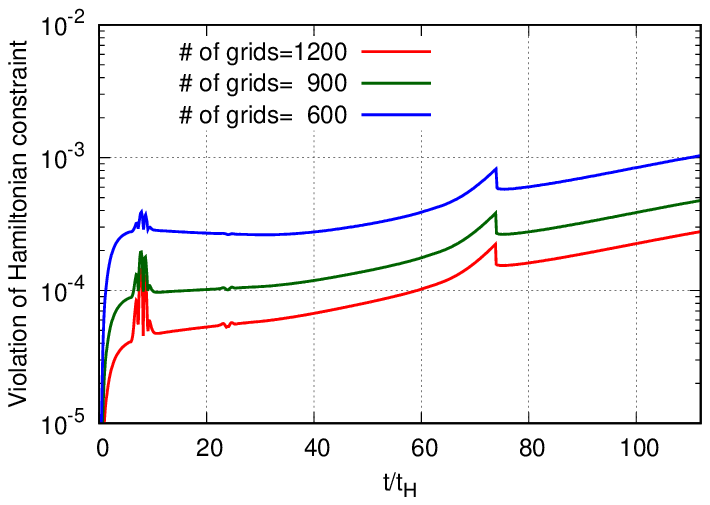}\caption{\baselineskip5mm
The average value of the violation of the Hamiltonian constraint with respect to 
the number of grid points as a function of time, for $\mu=0.62$ (left) 
and $\mu=0.64$ (right). 
For the $\mu=0.64$ case, the average is taken outside the black hole horizon after the horizon formation at around $t=74 t_H$. 
}
\label{fig:const}
\end{center}
\end{figure}

Let us closely look at the time evolution of the energy density 
around the horizon entry. 
In Fig.~\ref{fig:eneden}, we show the time evolution of the ratio 
of the scalar field energy density $E^{\rm sc}$ to 
the fluid energy density $E^{\rm f}$ (left panel) and the ratio of the
 total energy density given by $(\dd M_{\rm MS}/\dd R_{\rm A}) / (4\pi R_{\rm A}^2)$ 
 with $R_{\rm A}$ being the areal radius 
to the background energy density $\rho_{\rm b}$ (right panel), for $\mu=0.64$. 
We do not show the $\mu=0.62$ case because the behaviors are similar to those for $\mu=0.64$ around the horizon entry. 
Around the horizon entry, the kinetic energy of the scalar field makes a considerable contribution 
in the central region, and it gradually decays from $t\sim 10 t_{\rm H}$. 
Thus the collapsing dynamics is dominated by the fluid in the late time. 
The total energy density initially grows together with the energy density of the scalar field 
and keeps growing after the horizon entry. 
It is worthy to note that, at the initial time $t=0.04t_{\rm H}$, as is shown
in Fig.~3, the amplitude of the total energy density perturbation represented by
the Misner-Sharp mass is tiny compared 
to the background energy density. This implies that the peak value of the 
corresponding compaction function is also tiny.
This is the specific character of the isocurvature perturbation, 
and it can contribute to the 
nontrivial inhomogeneous geometry only around and after the horizon entry.  
%%%%%%%%%%%%%%%%%%%%%%%%%%%<<start figure>>%%%%%%%%%%%%%%%%%%%%%%%%%%
%%%%%%%%%%%%%%%%%%%%%%%%%%%%<<end figure>>%%%%%%%%%%%%%%%%%%%%%%%%%%%

In order to show the convergence of the simulation, 
we also plot the average value of the violation of the Hamiltonian
 constraint in Fig.~\ref{fig:const}. 
We note that the average is taken outside the black hole horizon if it exists. 
We find that the convergence is roughly quadratic in the spatial resolution.
Since we are using a second order scheme in the evaluation of the fluid flux, 
the order of the convergence is consistent with our procedure. 
%%%%%%%%%%%%%%%%%%%%%%%%%%%<<start figure>>%%%%%%%%%%%%%%%%%%%%%%%%%%
%%%%%%%%%%%%%%%%%%%%%%%%%%%%<<end figure>>%%%%%%%%%%%%%%%%%%%%%%%%%%%

%%%%%%%%%%%%%%%%%%%%%%%%%%%%%%%%%%%%%%%%%%%%%%%%%%%%%%%%%%%%%%%%
\subsection{Time evolution of the PBH mass}
\label{sec:fitmass}
%%%%%%%%%%%%%%%%%%%%%%%%%%%%%%%%%%%%%%%%%%%%%%%%%%%%%%%%%%%%%%%%
Let us check the time evolution of the mass of the PBH. 
Since the energy density of the scalar field is negligible compared to the fluid energy density during and after the collapse, 
we may expect that the black hole mass obeys the following simple formula firstly proposed by Zel'dovich and Novikov~\cite{1967SvA....10..602Z}~(see also Refs.~\cite{Custodio:1998pv,Deng:2016vzb} for its extension and numerical fitting):
\begin{equation}
\frac{\dd M}{\dd t}=16\pi F M^2\rho_{\rm b}, 
\end{equation} 
where $\rho_{\rm b}$ is the background energy density and $F$ is the constant which determines the accretion efficiency. 
Integrating this equation for the radiation-dominated universe, 
we obtain the following expression:
\begin{equation}
M(t)=\frac{1}{\frac{1}{M_{\rm f}}+\frac{3F}{2t}}, 
\label{eq:fiteq}
\end{equation}
where $M_{\rm f}$ is the final mass of the black hole. 
This formula is expected to be valid in a sufficiently late time 
after the black hole formation. 
There are some arguments about the value of 
$F$~\cite{1967SvA....10..602Z, Custodio:1998pv}.
When we compare the formula with a numerical result, 
the foliation of the time slice should be properly taken into account
(see Appendix for a simple discussion). 
In this paper, avoiding all these complexities, 
we simply treat the constant $F$ as a fitting parameter together with $M_{\rm f}$. 

\begin{figure}[htbp]
\begin{center}
\includegraphics[scale=1.6]{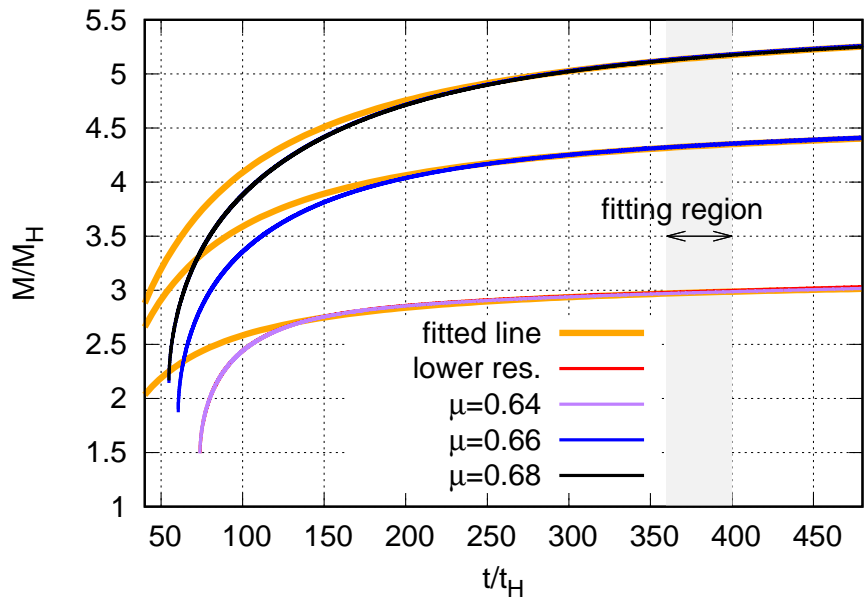}
\caption{\baselineskip5mm
The time evolution of the black hole mass is shown for each value of $\mu$. 
The yellow thick lines are fitted lines with the equation \eqref{eq:fiteq} which has two parameters $M_{\rm f}$ and $F$. 
The values of these fitting parameters are summarized in Table~\ref{tab:paras}. 
The fitting is performed in the shaded region in the graph. 
}
\label{fig:fitmass}
\end{center}
\end{figure}

We performed numerical simulations for the 3 cases: $\mu=0.64$, 0.66 and 0.68. 
The time evolution of the black hole mass is shown in Fig.~\ref{fig:fitmass}. 
We took 800 and 1000 grid points for the lower and higher resolutions, respectively. 
Since the results for the lower resolution almost 
overlap the results of the higher resolution, 
we conclude that the results are convergent. 
The fitting line for each result is shown as the yellow thick line. 
The fitting is performed only within a finite time interval
 shown in Fig.~\ref{fig:fitmass}. 
Even though the fitting is performed in the limited region, 
the formula fits well with the numerical result beyond the fitting region. 
This result implies that the time evolution of the PBH generated from the 
isocurvature perturbation can be well described by that in the case of
the purely radiation-dominated universe. 
The values of the fitting parameters are summarized in Table~\ref{tab:paras}. 
The values of $F$ are clearly larger than those reported in Refs.~\cite{Deng:2016vzb,Escriva:2019nsa}. 
This discrepancy is perhaps due to the difference 
in the foliation of the time slices~\cite{priv_Escriva}. 
A simple discussion about the time slice dependence of $F$ is given in Appendix.
%%%%%%%%%%%%%%%%%%%%%%%%%%%<<start figure>>%%%%%%%%%%%%%%%%%%%%%%%%%%
%%%%%%%%%%%%%%%%%%%%%%%%%%%%<<end figure>>%%%%%%%%%%%%%%%%%%%%%%%%%%%
%%%%%%%%%%%%%%%%%%%%%%%%%%%<<start table>>%%%%%%%%%%%%%%%%%%%%%%%%%%
\begin{table}[htbp]
\caption{The values of $F$ and $M_{\rm f}$ for each value of $\mu$. 
}
\label{tab:paras}
\begin{tabularx}{100mm}{C|CC}
$\mu$&$M_{\rm f}[M_{\rm H}]$&$F$\\
\hline
\hline
0.64&3.15&4.65\\
0.66&4.68&4.32\\
0.68&5.68&4.56\\
\hline
\end{tabularx}
\end{table}
%%%%%%%%%%%%%%%%%%%%%%%%%%%<<end table>>%%%%%%%%%%%%%%%%%%%%%%%%%%

%%%%%%%%%%%%%%%%%%%%%%%%%%%%%%%%%%%%%%%%%%%%%%%%%%%%%%%%%%%%%%%%
\subsection{Behavior near the threshold}
\label{sec:critical}
%%%%%%%%%%%%%%%%%%%%%%%%%%%%%%%%%%%%%%%%%%%%%%%%%%%%%%%%%%%%%%%%
As is shown in the previous sections, the contribution of the massless 
scalar field is negligible at the time of the horizon formation. 
Therefore we expect that the mass of the black hole obeys the critical 
mass scaling~\cite{Choptuik:1992jv,Koike:1995jm,Niemeyer:1997mt,Yokoyama:1998xd,Green:1999xm,Kuhnel:2015vtw,Niemeyer:1999ak,Musco:2004ak,Musco:2008hv,Musco:2012au} with the critical exponent $\gamma\simeq0.3558$\cite{Koike:1995jm}.\footnote{One should keep in mind that 
the critical scaling analysis can be done only under the assumption of 
spherical symmetry. Whether a similar scaling holds in a realistic cosmological 
situation is an unresolved issue.}
It is known that, in order to find the critical exponent in the PBH formation, 
we need to resolve the horizon formation over more than a few orders of magnitude 
in the resultant black hole mass. 
In order to achieve this resolution, we implemented fixed mesh refinement around
 the center. 
We introduced at most 8 layers depending on the resolution requirement
 for resolving the black hole formation. 

\begin{figure}[htbp]
\begin{center}
\includegraphics[scale=1.4]{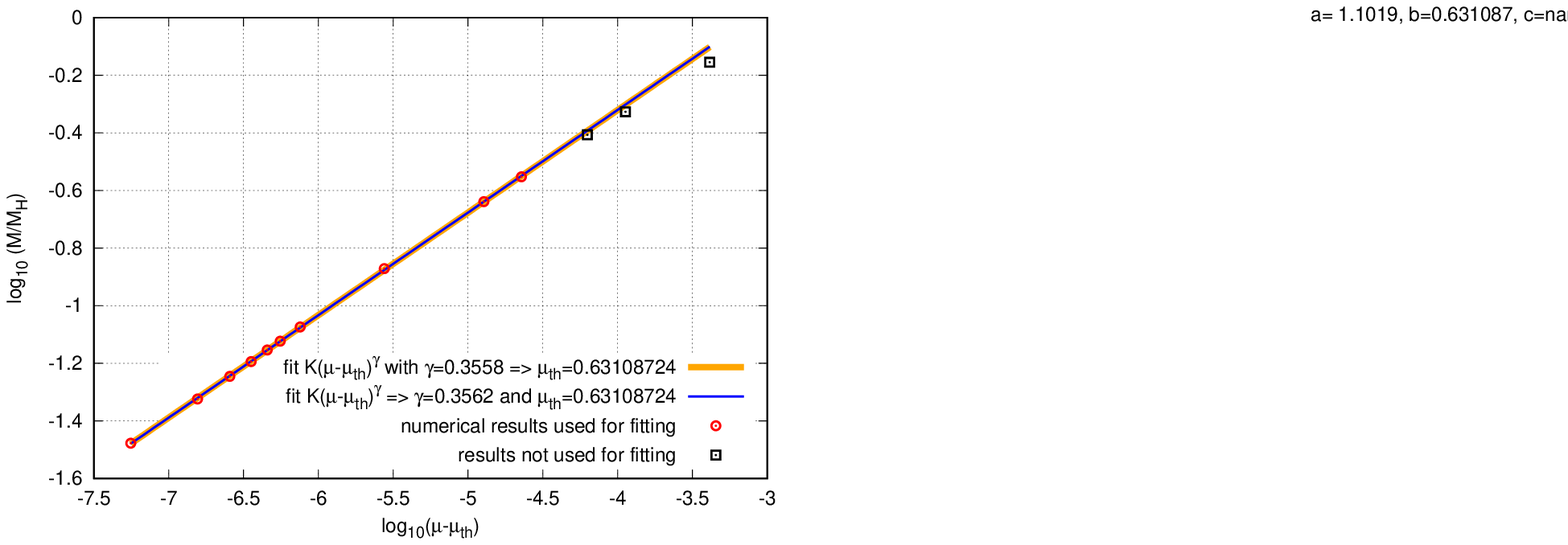}
\caption{\baselineskip5mm
The mass of the primordial black hole at the formation time is plotted as a function of the 
initial amplitude of the massless scalar field. 
The solid blue line is the fitted line of the functional form $M/M_H= \mathcal K(\mu-\mu_{\rm th})^\gamma$, 
and the best fit value of $\gamma$ is given by $0.3562$. 
The thick orange line is the fitted line with $\gamma$ fixed to be $0.3558$. 
For the fitting, we used the lower 10 points indicated by the red circle where 
the scaling relation seems applicable. 
We obtain $\mu_{\rm th}\simeq0.63108724$ in the fitting for both cases, and 
$\mathcal K\simeq12.6$ and $12.7$ for the orange and blue lines, respectively. 
}
\label{fig:critical}
\end{center}
\end{figure}

In Fig.~\ref{fig:critical}, we show the mass of the black hole at the moment of the horizon formation 
as a function of the initial amplitude of the massless scalar field. 
We performed the fitting by using lower 10 data points shown in Fig.~\ref{fig:critical} 
where the scaling relation seems to be realized. 
The thick orange line in Fig.~\ref{fig:critical} shows the line with $\gamma=0.3558$ while 
the blue solid line is the fitted line without assuming the value of the exponent $\gamma$, 
where the best fit value of $\gamma$ is given by $0.3562$. 
The threshold value given by the fitting is $\mu_{\rm th}\simeq0.63108724$. 
Our results show good agreement with the critical scaling in the radiation dominated universe. 
%%%%%%%%%%%%%%%%%%%%%%%%%%%<<start figure>>%%%%%%%%%%%%%%%%%%%%%%%%%%
%%%%%%%%%%%%%%%%%%%%%%%%%%%%<<end figure>>%%%%%%%%%%%%%%%%%%%%%%%%%%%

%%%%%%%%%%%%%%%%%%%%%%%%%%%%%%%%%%%%%%%%%%%%%%%%%%%%%%%%%%%%%%%%
\section{Summary and discussion}
\label{sec:sumdis}
%%%%%%%%%%%%%%%%%%%%%%%%%%%%%%%%%%%%%%%%%%%%%%%%%%%%%%%%%%%%%%%%
We studied the PBH formation from an isocurvature perturbation 
of a massless scalar field in the radiation-dominated universe. 
Deriving the general growing mode solution of the isocurvature perturbation 
to second order in spatial gradient expansion, 
we performed numerical simulations of the PBH formation starting
 from a spherically symmetric Gaussian spatial profile of the scalar field.
Even though there is no contribution of the massless scalar field to 
the stress-energy tensor at leading order, the growing mode solution 
generally contains the non-trivial metric perturbation at second order,
which may consequently lead to non-linear dynamics of the geometry
after the horizon entry. 
We actually found that a PBH can form for a sufficiently large initial amplitude 
of the isocurvature perturbation. 

We found that although the PBH formation is originally caused by the scalar field 
isocurvature perturbation, the radiation fluid dominates the dynamics 
during the gravitational collapse. As a result, the evolution of the PBH mass 
due to the accretion and the critical behavior near the threshold mass
show similar behaviors to those in the standard PBH formation 
in the radiation-dominated universe. 
These facts imply that all the constraints on the curvature perturbation 
through the observational upper limits of PBH abundance 
can apply to the case of the isocurvature perturbation of the massless scalar field, 
and we can obtain similar observational constraints on it. 

In this paper, we considered the radiation fluid and a massless scalar field 
for the matter components as a first step of various possibilities. 
It would be interesting to consider a scalar field with a non-trivial 
potential term which introduces additional physical scales, 
and a rich variety of dynamics across the horizon entry may 
be expected~\cite{deJong:2021bbo}. 
The simulation of the PBH formation from the dark matter 
isocurvature~\cite{Passaglia:2021jla}, 
for which non-spherical dynamics may be essential, 
would be also interesting although such a simulation 
is much more challenging as one cannot assume spherical symmetry. 
There are many other fascinating directions to be explored. 
The current work is just the beginning of these unexplored research themes.

%%%%%%%%%%%%%%%%%%%%%%%%%%%%%%%%%%%%%%%%%%%%%%%%%%%%%%%%%%%%%%%%
\section*{Acknowledgements}
%%%%%%%%%%%%%%%%%%%%%%%%%%%%%%%%%%%%%%%%%%%%%%%%%%%%%%%%%%%%%%%%
We thank Albert Escriv\`a for helpful comments and stimulating discussion. 
This work was supported in part by JSPS KAKENHI Grant Nos.
JP19H01895 (CY, TH, SH, MS), JP20H05850 (CY), JP19K03876 (TH), JP20H04727 (MS) 
and JP20H05853 (CY, TH, SH, MS). 

\appendix

\section{Time slice dependence of $F$}

Let us introduce the effective time $t_{\rm h}$ near the horizon as follow
\begin{equation}
t_{\rm h}=\alpha_{\rm h} t_{\rm c},  
\end{equation}
here and hereafter, the subscripts ``$~_{\rm h}$" and ``$~_{\rm c}$" represent the variables for the near horizon region 
and the cosmological background, respectively. 
$\alpha_{\rm h}$ is the lapse function near the horizon and $\alpha_{\rm c}$ is set to 1.  
We consider $\alpha_{\rm h}$ as a constant for simplicity. 
The density scales as $\rho\propto t^{-2}$, therefore we can define the effective density 
near the horizon as 
\begin{equation}
\rho_{\rm h}=\alpha_{\rm h}^{-2}\rho_{\rm c}. 
\end{equation}
Then, considering the differential equation of the black hole mass around the horizon, 
\begin{equation}
\frac{dM}{dt_{\rm h}}=4\pi F_{\rm h} (2M)^2\rho_{\rm h}, 
\end{equation}
we obtain 
\begin{equation}
%\Leftrightarrow
\frac{dM}{dt_{\rm c}}=4\pi \frac{F_{\rm h}}{\alpha}(2M)^2\rho_{\rm c}
\end{equation}
Therefore the effective coefficient $F_{\rm c}$ is given by 
\begin{equation}
F_{\rm c}=\frac{F_{\rm h}}{\alpha}. 
\end{equation}
Since what we see by the fitting is the value of $F_{\rm c}$, 
the simple analysis suggests the inverse correlation between the lapse function near the horizon and the 
fitting parameter $F$.

% \bibliographystyle{h-physrev5-title}
% \bibliography{../bibfiles/pbh}

\end{document}